\PassOptionsToPackage{prologue,table}{xcolor}
\documentclass[sigconf]{acmart}

\usepackage{amsmath,amsfonts}
\usepackage{algorithmic}
\usepackage{graphicx}
\usepackage{textcomp}
\usepackage[table]{xcolor}
\usepackage{tcolorbox}
\usepackage{hyperref}
\usepackage{booktabs}
\usepackage{multirow}
\usepackage[flushleft]{threeparttable}
\usepackage{soul}
\usepackage{natbib}
\setcitestyle{square, comma, numbers,sort&compress, super}
\usepackage{twemojis}
\usepackage[applemac]{inputenc}
\usepackage{makecell}
\usepackage{balance}
\usepackage{pifont}
\usepackage{listings}
\usepackage{xcolor}
\usepackage{xspace}

\definecolor{Large}{HTML}{696969}
\definecolor{Negligible}{HTML}{D3D3D3}
\definecolor{Medium}{HTML}{808080}
\definecolor{Small}{HTML}{A9A9A9}
\usepackage{enumitem}
\usepackage{caption}

\AtBeginDocument{%
  \providecommand\BibTeX{{%
    \normalfont B\kern-0.5em{\scshape i\kern-0.25em b}\kern-0.8em\TeX}}}

\copyrightyear{2024}
\acmYear{2024}
\setcopyright{acmlicensed}
\acmConference[ICSE '24]{2024 IEEE/ACM 46th International Conference on Software Engineering}{April 14--20, 2024}{Lisbon, Portugal}
\acmBooktitle{2024 IEEE/ACM 46th International Conference on Software Engineering (ICSE '24), April 14--20, 2024, Lisbon, Portugal}
\acmDOI{10.1145/3597503.3639127}
\acmISBN{979-8-4007-0217-4/24/04}

\newcommand{\RqOne}{\textbf{RQ1:} \emph{How are \GHS~profiles discussed on Twitter/X?}}

\newcommand{\RqOneDotOne}{\emph{
What are the characteristics of tweets mentioning GitHub Sponsors profiles from organizational and personal accounts?
}}

\newcommand{\RqOneDotTwo}{\emph{Who mentions \GHS~profiles on Twitter/X?}\nopunct}

\newcommand{\RqOneDotThree}{\emph{What is the context of \GHS~profile mentions on Twitter/X?}\nopunct}

\newcommand{\RqOneDotFour}{\emph{When are \GHS~profiles mentioned
on Twitter/X?}\nopunct}

\newcommand{\RqTwo}{\textbf{RQ2:} \emph{What is the impact of \GHS~profile mentions on Twitter/X?}}

\newcommand{\RqTwoDotOne}{\emph{How are \GHS~profile mentions received on Twitter/X?}\nopunct}

\newcommand{\RqTwoDotTwo}{\emph{How are \GHS~profile mentions discussed on Twitter/X?}\nopunct}

\newcommand{\RqTwoDotThree}{\emph{How do \GHS~profile mentions impact sponsorship?}\nopunct}

\newcommand{\GHS}{GitHub Sponsors}

\newcommand{\revise}[1]{\textcolor{black}{#1}}

\graphicspath{{figs}}

\begin{document}

\title{``My \GHS~profile is live!''  Investigating the Impact of Twitter/X Mentions on \GHS}


\author{Youmei Fan}
\affiliation{%
  \country{Nara Institute of Science and Technology, Japan}
}
\email{fan.youmei.fs2@is.naist.jp}
\author{Tao Xiao}
\affiliation{%
  \country{Nara Institute of Science and Technology, Japan}
}
\email{tao.xiao.ts2@is.naist.jp}
\author{Hideaki Hata}
\affiliation{%
  \institution{Shinshu University}
  \country{Japan}
}
\email{hata@shinshu-u.ac.jp}
\author{Christoph Treude}
\affiliation{%
  \institution{University of Melbourne}
  \country{Australia}
}
\email{christoph.treude@unimelb.edu.au}
\author{Kenichi Matsumoto}
\affiliation{%
  \country{Nara Institute of Science and Technology, Japan}
}
\email{matumoto@is.naist.jp}


\begin{abstract}
\GHS~was launched in 2019, enabling donations to open-source software developers to provide financial support, as per GitHub's slogan: ``Invest in the projects you depend on''. However, a 2022 study on \GHS~found that only two-fifths of developers who were seeking sponsorship received a donation. The study found that, other than internal actions (such as offering perks to sponsors), developers had advertised their \GHS~profiles on social media, such as Twitter (also known as X). Therefore, in this work, we investigate the impact of tweets that contain links to \GHS~profiles on sponsorship, as well as their reception on Twitter/X. We further characterize these tweets to understand their context and find that (1) such tweets have the impact of increasing the number of sponsors acquired, (2) compared to other donation platforms such as Open Collective and Patreon, \GHS~has significantly fewer interactions but is more visible on Twitter/X, and (3) developers tend to contribute more to open-source software during the week of posting such tweets. Our findings are the first step toward investigating the impact of social media on obtaining funding to sustain open-source software.
\end{abstract}



\keywords{Open-source Software, Sponsorship, Social Media}



\maketitle

\section{Introduction}
Open-source software (OSS) is ubiquitous, but sustaining it is a challenge~\cite{10.1145/3377811.3380410}. Maintaining an OSS project requires not only intrinsic motivation (e.g., joy of participation) but also extrinsic motivation (e.g., financial incentives)~\cite{zhou2022studying}. The last few years have seen the emergence of many platforms that allow open-source developers to receive donations for their work, such as
PayPal~\cite{paypalDigitalWallets},
Open Collective~\cite{opencollectiveRaiseSpend},
Patreon~\cite{patreonCreativityPowered},
and GitHub Sponsors~\cite{githubGitHubSponsors}.
Several platforms support sponsoring OSS projects in cryptocurrencies, with the rise in popularity of cryptocurrencies today, e.g., Gitcoin Grants~\cite{gitcoinGrants}
and Giveth~\cite{givethGivethWelcome}.
However, as Overney et al.'s paper title ``How to not get rich: an empirical study of donations in open source''~\cite{10.1145/3377811.3380410} suggests, simply having a platform for donations is not enough. In a 2022 study on \GHS, Shimada et al.~\cite{shimada2022github} found that out of approximately 9,000 developers who had activated their \GHS~profile, less than 40\% had received a donation. 

If simply creating a sponsorship profile is not enough, what else can open-source software developers do to attract donations? Following the long line of work on studying the intersection between social media and software development~\cite{storey2016social, storey2010impact, fang2022damn}, in this paper, we investigate the impact of tweeting about a \GHS~profile on sponsorship. To make it easy for its users to reach a large audience, GitHub provides tweet templates that users can use to advertise a new \GHS~profile (``My \GHS~profile is live! You can sponsor me to support my open source work \twemoji[width=1em]{sparkling heart}'') or to broadcast that they made a donation (``\twemoji[width=1em]{sparkling heart} I'm sponsoring [username] because...'').
The impact of tweets on open-source software development has been investigated before. Fang et al.~\cite{fang2022damn} found that tweets have a significant effect on obtaining new stars and new contributors for an open-source project and that the formation of an active Twitter/X community plays an important role in attracting new contributors. The role of tweets has also been studied in the context of bug fixing~\cite{mezouar2018tweets} and trend awareness~\cite{singer2014software}. To the best of our knowledge, the role of Twitter/X in obtaining funding for open-source development has not yet been studied.

We first characterize the state of the practice by quantitatively and qualitatively analyzing more than 10,000 tweets linking to \GHS~profiles to understand the context of such tweets. We then measure the impact of the tweets in terms of their reception on Twitter/X and their effect on sponsorship.
We found links to \GHS~profiles on Twitter/X are common and the majority of such tweets are written by users other than the profile owner, such as a sponsor. We identified a significant positive effect of GitHub Sponsor profile mentions in tweets on the number of sponsors acquired.
Tweet mentions have the impact of increasing the number of sponsors by 1.22. Although \GHS~has surpassed other platforms such as Open Collective and Patreon in terms of visibility on Twitter/X, tweets about \GHS~received significantly fewer likes, retweets, and replies compared to other platforms. Developers tended to be more active during the week of a tweet, in particular, in terms of the number of commits. 

\textbf{Significance of research contribution.}
The findings of our study have significant implications, indicating a strong interconnection between social media channels and donation pathways within the social programmer ecosystem~\cite{treude2012programming}. Our research demonstrates that actively engaging on social media platforms to promote sponsorship opportunities for open-source development can yield fruitful outcomes. This suggests that open-source developers stand to benefit from expanding their presence and networking efforts beyond the GitHub platform. Furthermore, our study highlights the notion that publicity and visibility in the realm of open-source sponsorship need not be limited to a unidirectional flow. Rather, sponsors themselves have the potential to enhance the exposure and reach of open-source projects by publicizing their donations. In doing so, they serve as exemplars, setting a positive precedent and inspiring others to follow suit. By emphasizing these key findings, we provide compelling evidence to support the notion that using social media channels, diversifying online networks, and fostering mutual publicity between sponsors and developers can yield substantial advantages within the open-source community. These insights encourage open-source developers and sponsors alike to consider the broader potential of social media engagement and collaborative promotion to achieve their goals.
\section{Research Questions}
%

The main objective of our study is to understand the state of practice and the impact of \GHS~profile mentions on Twitter/X. The insights drawn from this study will not only contribute to the academic understanding but also have practical implications for developers, sponsors, and platforms like GitHub. Furthermore, our findings can shed light on the relationship between social activities and monetary contributions, ultimately serving to augment the appeal of developers. To guide our investigation, we present main questions and sub-questions, along with motivations and relevance.

\noindent
\RqOne 
~The motivation behind RQ1 is to provide insights into the dynamics of \GHS~profile mentions, ultimately informing better strategies for developers seeking sponsorship.\\
\indent \textbf{RQ1.1} \RqOneDotOne 
~Understanding the language, account types, and programming languages in these tweets will enable developers to craft more appealing content for potential sponsors, ultimately enhancing engagement.\\
\indent \textbf{RQ1.2} \RqOneDotTwo 
~By identifying who engages with these profiles, sponsorship acquisition strategies can be tailored to target specific demographics.\\
\indent \textbf{RQ1.3} \RqOneDotThree 
~Investigating the context in which profiles are mentioned will shed light on why tweets are used in sponsorship communication, potentially informing strategies for developers seeking sponsorship.\\
\indent \textbf{RQ1.4} \RqOneDotFour 
~Analyzing the timing of mentions can lead to the discovery of optimal moments for posting, which could help in securing sponsorship.


\noindent
\RqTwo 
~Building on RQ1, RQ2 explores the effects of the dynamics uncovered, allowing us to measure and interpret their impact.
\indent \textbf{RQ2.1} \RqTwoDotOne 
~Understanding the reception will aid platforms like GitHub in providing targeted support and tools to developers, such as social media templates and guidelines.\\
\indent \textbf{RQ2.2} \RqTwoDotTwo 
~By examining engagement metrics and replies, we will gain a deeper understanding of the conversations, ultimately helping in crafting more effective strategies for community engagement.\\
\indent \textbf{RQ2.3} \RqTwoDotThree 
~Through a quasi-experimental approach, our goal is to provide quantitative evidence of the causal impact, which can guide both developers in improving their social media strategies and platforms in enhancing features that facilitate sponsorship acquisition.

By addressing these research questions, we aim to provide insights into the dynamics and consequences of \GHS~mentions on Twitter/X. This exploration contributes to theoretical understanding and practical strategies, offering value to the broader Open Source community.



\section{Research Methods}

This section describes our methods for data collection and our quantitative and qualitative analyses.

\subsection{Data Collection}
\label{ssec:dc}
In this study, we examine tweets containing links to \GHS~profiles.
We successfully applied for Twitter/X's Academic Research Access~\cite{twitterTwitterAcademic},
which offers a higher limit on the number of tweets that can be retrieved per month, and we analyzed tweets from May 2019, when \GHS~was launched, through April 2022, using \texttt{Twitter/X API v2}~\cite{twitterTwitterDocumentation}
in May 2022. The Twitter/X API provides a search function that allows for a set of query mechanisms against tweets. We use the ``url'' query to retrieve tweets that contain links with the specific substring ``\texttt{github.com/}\\ \texttt{sponsors/}'' so that we ensure all the tweets are developer-related. We obtained 11,582 tweets that contain \GHS~profile links.
Among these tweets, the majority (91\%) were written in English, accounting for 10,531 tweets. We only use English tweets for the following quantitative and qualitative analyses, except \textbf{RQ1.1}.


\subsection{Quantitative Analysis}
\label{ssec:quantitative}

To understand the characteristics of tweets mentioning \GHS~profiles (\textbf{RQ1.1}),
we investigate written languages, types of GitHub accounts, and primary programming languages of developers mentioned in the tweets. For written languages, we calculate the distribution of languages in tweets. In cases where Twitter/X cannot determine the language of a tweet (e.g., the tweet only contains hashtags, emojis, or links), \texttt{Undetermined} is used. 

Since one \GHS~profile may appear in different tweets, we obtained distinct \GHS~profiles in tweets to collect the types of GitHub accounts and the primary programming languages of the corresponding developers. We obtained 3,766 distinct \GHS~profiles from the 11,582 tweets. For the types of GitHub accounts, we calculate the distribution of the types of GitHub accounts (i.e., personal or organizational) across all distinct \GHS~profiles in tweets. Since the URL of a \GHS~profile is organized as \texttt{https://github.com/sponsors/} \texttt{[username]}, we can retrieve the corresponding GitHub account using \texttt{username} in the GitHub GraphQL API~\cite{githubGitHubGraphQL}.

The primary programming languages of the repositories can also be retrieved using the GitHub GraphQL API.
Same to the previous work~\cite{shimada2022github}, we take the most common primary language of the repositories to which each developer contributed as the primary language of that developer. This
is an approximation because we did not analyze whether the developer actually committed in that language. If the occurrences of each programming language per repository are the same, we consider the primary programming language to be \texttt{Undetermined}. The primary languages of developers identified in this way can be interpreted as the programming languages of the ecosystems to which the developers mainly contributed.

To attract potential sponsors, developers might be particularly active on GitHub around the time they advertise their \GHS~profile on Twitter/X. To investigate whether such correlations exist, we considered three time periods related to a ``My
\GHS~profile is live!'' tweet, i.e., a week before posting this tweet, the week in which the tweet was posted, and a week after posting this tweet (\textbf{RQ1.4}). For example, if a tweet has been posted on 15 June 2020, these three periods will be from \texttt{2020-06-05} to \texttt{2020-06-11}, \texttt{2020-06-12} to \texttt{2020-06-18}, and \texttt{2020-06-19} to \texttt{2020-06-25}, respectively. Following the approach of related work~\cite{capiluppi2013effort}, which used a time frame of one week before and after, our decision to adopt a one-week duration allows us to quickly assess immediate changes in productivity, engagement, and quality. This analysis involves scrutinizing short-term developer activities before and after sponsorship, facilitating a timely evaluation.
We obtained 810 distinct \GHS~profiles that were posted using that template from our data set.
Then we investigate different categories of contribution activities in each period, as shown below. To collect these contribution activities in a week, we retrieve them from the profile pages of the GitHub accounts as \texttt{https://github.com/[username]} \texttt{?tab=overview\& from=[time period]\&to=[time period]}. 

\begin{itemize}[topsep=5pt]
    \item \revise{\textbf{opening pull request}: The profile page indicates that the GitHub account has opened pull requests, including substrings such as ``\textit{Created a pull request}'',
``\textit{Opened 1 other pull request}'',
``\textit{Opened [number] pull requests}'', and
``\textit{Opened their first pull request}''.
    \item \textbf{submitting pull request review}: The profile page indicates that the GitHub account has reviewed pull requests, including a substring such as ``\textit{Reviewed [number] pull requests}''.
    \item \textbf{opening issue}: The profile page indicates that the GitHub account has opened issues, including substrings such as ``\textit{Created an issue}'',
``\textit{Opened [number] other issues}'',
``\textit{Opened their first issue}'', and
``\textit{Opened [number] issues}''.
    \item \textbf{opening discussion}: The profile page indicates that the GitHub account started a GitHub Discussion, including a substring such as ``\textit{Started [number] discussions}''.
    \item \textbf{answering discussion}: The profile page indicates that the GitHub account answered a GitHub Discussion, including a substring such as ``\textit{Answered [number] discussions}''.
    \item \textbf{committing}: The profile page indicates that the GitHub account has authored commits, including a substring such as ``\textit{Created [number] commits}''.
    \item \textbf{contributing in private repository}: The profile page indicates that the GitHub account contributed to private repositories, including a substring such as ``\textit{[number] contributions in private repositories}''.
    \item \textbf{creating repository}: The profile page indicates that the GitHub account created private repositories, including substrings such as ``\textit{Created [number] other repositories}'', 
``\textit{Created [number] repositories}'', and
``\textit{Created their first repository}''.
\item \textbf{joining organization}: The profile page indicates that the GitHub account joined an organization, including a substring such as ``\textit{Joined the [name] organization}''.}
\end{itemize}
 We conduct Mann-Whitney U tests to compare activities in these three periods, i.e., between a week before posting the tweet and the week when the tweet was posted, and between the week when the tweet was posted and a week after posting this tweet.
To estimate the effect size of significant differences, we use Cliff's delta with the following thresholds~\cite{romano2006appropriate}: negligible for 0 $\leq$ ~$|$$\textit{delta}$$|$ $<$ 0.147, small for 0.147 $\leq$ ~$|$$\textit{delta}$$|$ $<$ 0.33, medium for 0.33 $\leq$ ~$|$$\textit{delta}$$|$ $<$ 0.474, and large otherwise. 

To investigate the reception of tweets mentioning \GHS~profiles (\textbf{RQ2.1}), we analyze the popularity of tweets that mentioned \GHS~profiles on Twitter/X (number of likes, number of retweets, and number of replies). Then, we compare these interactions to tweets that contain links to other donation and crowd-funding platforms that are often used to obtain financial support for OSS development~\cite{10.1145/3377811.3380410}:
PayPal, 
Open Collective, 
and Patreon. 
 To ensure that the tweets obtained are related to OSS, we collect tweets that contain links to at least one of these three platforms and GitHub (i.e., ``\texttt{github.com}'', except links to \GHS) using \texttt{Twitter/X API v2} in the same time period for which we collected \GHS~profile tweets. We consider a link to point to a PayPal profile when it contains ``\texttt{paypal.com/paypalme/}'', Open Collective when it contains ``\texttt{opencollective.com/}'', and Patreon when it contains ``\texttt{patreon.com/}'', except Patreon posts (i.e., ``\texttt{patreon.com/posts/}''). We exclude tweets that contain links to ``\texttt{github.com/sponsors/}'' and at least one of these three platforms from the 10,531 English tweets obtained. In the end, we obtained 10,440 tweets for \GHS, four tweets for PayPal, 88 tweets for Open Collective, and 228 tweets for Patreon. Since only four tweets contain links to PayPal, we focus on comparisons between \GHS, Open Collective, and Patreon. 

We also conduct Mann-Whitney U tests to compare Twitter/X
interactions between \GHS~and Open Collective, and between \GHS~and Patreon.
We use Cliff's delta with the same thresholds to estimate the effect size of significant differences.

\subsection{Qualitative Analysis}
\label{ssec:qualitative}

For our qualitative analyses, we randomly selected a statistically representative number of tweets with a confidence level of 95\% and a confidence interval of 5 to obtain 371 tweets from the initial population of 10,531 English-language tweets. 

Unsurprisingly, our initial analysis revealed differences between the dynamics around tweets from users looking for sponsorship and those from users who made donations. Therefore, we categorized the 371 tweets into three different behavioral groups based on the purpose of the tweet, so we could see what kind of tweets a user would make based on their behavior:
\begin{itemize}
    \item \textbf{looking for sponsors}: This tweet is posted by developers who publicized their profiles to look for sponsors, e.g.,``\textit{My GitHub Sponsors profile is live! You can sponsor me to support my open source work \twemoji[width=1em]{sparkling heart}}''.
    \item \textbf{sponsors}: This tweet is posted by developers who sponsored others, e.g., ``\textit{\twemoji[width=1em]{sparkling heart}I'm sponsoring [username] because...}''.
    \item \textbf{no purpose}: This tweet does not have sufficient information to decide, e.g., ``\textit{You guys make magic.}''.
\end{itemize}
In the end, we identified 183 tweets from developers that were looking for sponsors, 168 tweets from sponsors, and 20 tweets from no purpose. These 351 tweets (20 tweets from no purpose are excluded) are used for answering RQ1.2--RQ1.4, and RQ2.2.

Four of the authors collaboratively took an initial look at a randomly selected subset of 30 tweets from the sample of 351 tweets, discussed which themes were present in the data and how these themes related to the research questions, and then formalized this discussion into coding schemata.
For each aspect that entailed manual coding, a total of 30 tweets were independently labeled by four annotators, resulting in Cohen's kappa exceeding 0.6 for all parts and even reaching 0.94 for \textbf{RQ2.2}.

Encouraged by the initial kappa agreements, the first two authors independently coded the remaining sample of 321 tweets. Then, they recalculated kappa agreements to assess the improvement in understanding of the coding schemata after labeling the first 30 tweets. Finally, four authors engaged in collaborative discussions to attain a consensus in cases of disagreement.
We attribute this stability to the fact that we had an initial discussion about all data, that tweets are relatively short, and that this particular team of authors has experience working together on qualitative data analysis from previous research projects.
We describe the coding schemata related to each research question in the following paragraphs.

To investigate who mentions \GHS~profiles on Twitter/X (\textbf{RQ1.2}),
we analyze the relationship between the authors of the tweets and the GitHub accounts that are linked in the tweets. Furthermore, the rationale behind having a ``user'' category in the aforementioned code is rooted in the goal of acquiring insights into what extent users benefit from a developer's project and are willing to voluntarily advertise the developer, thereby enabling the developer to obtain more sponsorship. The prevalence of the code ``non-specific'' in the results indicates that some users advertise for others without a specific purpose.
Since the names of accounts on Twitter/X and GitHub do not necessarily have to be the same, we employed qualitative analysis for this investigation. It is important to highlight that our decision not to employ automated techniques for verifying the association between Twitter/X and GitHub accounts was motivated by the realization that these techniques often fail to account for certain scenarios. For example, when a GitHub account is classified as an organization type and one of its members posts a tweet, it should be regarded as emanating from the same user. Consequently, we opt for a cautious approach, acknowledging the limitations of automated techniques and acknowledging the need for context-sensitive judgment in determining the correspondence between Twitter/X and GitHub accounts.
The four annotators independently labeled 30 tweets. Then, we calculate the kappa agreement of our coding schemata from four annotators. The initial Cohen's kappa for this qualitative analysis is 0.75, which
indicates `substantial' agreement~\cite{viera2005understanding}. For the remaining sample of 321 tweets, Cohen's kappa is 0.78, which also indicates `substantial' agreement, from the first two authors. 
Examples of the following codes are covered in our replication package, aiming to facilitate the reader's comprehension of this taxonomy.

\begin{itemize}
   \item \textbf{same}: The author of this tweet is the same as the GitHub account that is shown on the \GHS~profile linked in the tweet, or the content of the tweet implies that they are the same developer or the author belongs to the GitHub organizational account on that \GHS~profile.

  \item \textbf{user}: The tweet explicitly indicates that the author of this tweet is a user of an open-source project that belongs to the GitHub account on the \GHS~profile.
  
  \item \textbf{non-specific}: There is not sufficient information to determine the relationship between the tweet author and the GitHub account.


\end{itemize}

To understand the context of tweets mentioning \GHS~profiles (\textbf{RQ1.3}),
we analyze why \GHS~profiles were mentioned in tweets. 
Additionally, the reason to distinguish between the ``advertisement with new information'' and ``advertisement with new functionality'' categories in the aforementioned coding schemata is to enable a more nuanced analysis: the former encompasses a range of updates, including changes to users' profile descriptions and tier information whereas the latter is related to functionality in the projects they are dedicated to. The four annotators independently
coded 30 tweets, achieving the initial Cohen's kappa of 0.66
or `substantial' agreement~\cite{viera2005understanding}. The first two authors then independently labeled the remaining sample of 321 tweets, finally reaching Cohen's kappa of 0.84 or `almost
perfect' agreement. The following list shows the coding schema that emerged from the data. Examples of the following codes are described in detail in our replication package to help the reader understand this taxonomy.
\begin{itemize}
     \item \textbf{generic advertisement}: This tweet advertises the tweet author's own \GHS~profile (use this code if the tweet does not fit the other advertisement categories).
     
     \item \textbf{donation appreciation}: This tweet explicitly expresses appreciation of a donation.

     \item \textbf{sponsor template}: This tweet contains GitHub's template for advertising one's own \GHS~profile: ``\textit{My GitHub Sponsors profile is live! You can sponsor me to support my open source work\twemoji[width=1em]{sparkling heart}}'' with no or minor changes.

     \item \textbf{advertisement of developer}: This tweet advertises the \GHS~profile of another personal GitHub account.

      \item \textbf{advertisement with new functionality}: This tweet explicitly advertises the author's own \GHS~profile while mentioning new functionality of an open-source project.
      
      \item \textbf{advertisement with new information}: This tweet explicitly advertises the author's own \GHS~profile with an update.

    \item \textbf{sustainability}: This tweet explicitly indicates an appreciation or need for a donation for the sustainability of an open-source project, often associated with terms such as ``sustainable''. 

      \item \textbf{advertisement with early access}: This tweet explicitly advertises the author's own \GHS~profile with early access to features (usually accompanied by a phrase such as ``early access'' and ``insider'')
    
    \item \textbf{income}: This tweet explicitly indicates the need for income to support one's daily life.

     \item \textbf{advertisement with event}: This tweet explicitly advertises a \GHS~profile with an event.

      \item \textbf{set example / peer pressure}: This tweet explicitly motivates others in either a positive or negative way.

    \item \textbf{advertisement of organization}: This tweet advertises the \GHS~profile of another organizational GitHub account.
   
    \item \textbf{donation to developer announcement}: This tweet explicitly indicates that the author of this tweet donated to the personal GitHub account in the \GHS~profile.
    
    \item \textbf{donation to organization announcement}: This tweet explicitly indicates that the author of this tweet donated to the organizational GitHub account in the \GHS~profile.
    
    \item \textbf{donation to developer template}: This tweet contains GitHub's template that indicates donation to a personal GitHub account: ``\textit{\twemoji[width=1em]{sparkling heart}I'm sponsoring [username] because...}'' with no or minor changes.
    
    \item \textbf{donation to organization template}: This tweet contains GitHub's template that indicates donation to an organizational GitHub account: ``\textit{\twemoji[width=1em]{sparkling heart}I'm sponsoring [username] because...}'' with no or minor changes.
    
\end{itemize}
To study when tweets related to \GHS~profiles occur in relation to other activities on GitHub (\textbf{RQ1.4}),
we analyze the timing of such tweets. The four annotators independently
coded 30 tweets, achieving the initial Cohen's kappa of 0.62
or `substantial' agreement~\cite{viera2005understanding}. For the remaining 321 tweets, the first two authors reached Cohen's kappa of 0.85 or `almost perfect' agreement. The following list shows the
coding schemata that emerged from the data. 
\begin{itemize}

    \item \textbf{start}: This tweet was posted when the \GHS~profile is activated (usually accompanied by a phrase like ``profile is live'').
    
    \item \textbf{no specific timing}: This tweet was posted with no particular timing.
    
    \item \textbf{donation}: This tweet was posted when the author of the tweet received a donation.

    \item \textbf{update}: This tweet was posted when there was an update to a GitHub project or \GHS~profile.
     \item \textbf{reach goal}: This tweet was posted in relation to reaching a goal.
     \item \textbf{release}: This tweet explicitly indicates that a release of the software project has been delivered.

    \item \textbf{event}: This tweet was posted when an event has been announced.
    \item \textbf{resignation / paycut}: This tweet was posted during a change in the author's work professional situation.

     \item \textbf{benefit}: This tweet explicitly mentions a particular benefit.
    \item \textbf{activity spike}: This tweet was posted to indicate the GitHub developer was particularly active and explicitly mentions the activity spike.
    
\end{itemize}

To investigate the responses to tweets mentioning \GHS~profiles (\textbf{RQ2.2}),
we analyze the replies to tweets mentioning \GHS~profiles. Four
annotators independently annotated 30 tweets. The initial kappa
agreement is 0.94, interpreted as `almost perfect' agreement~\cite{viera2005understanding}. The first two authors independently annotated the remaining 321 tweets, reaching the same kappa agreement. Our coding schemata emerged from the data and is as follows.
Note that examples for these codes are described in detail in our replication package to help the reader understand this taxonomy.
\begin{itemize}
    \item \textbf{support}: The response to this tweet demonstrates endorsement or encouragement for the author, often extending beyond appreciation and indicating a willingness to assist or advocate for the author's cause.
    
    \item \textbf{appreciation of work}: The respondent acknowledges and values the author's open-source contributions and their impact, without necessarily conveying explicit support or a commitment to assist in further efforts.

    \item \textbf{appreciation of donation}: The respondent to this tweet appreciates the donation.
    
    \item \textbf{emoji only}: The response to this tweet only contains emoji.
    \item \textbf{other}: The response to this tweet does not fit into the categories above, or there is no response to this tweet.

\end{itemize}

\subsection{Causal Inference}

We conduct a quasi-experiment to estimate the causal impact of \GHS~profile mentions in tweets on the number of sponsors acquired (\textbf{RQ2.3}).
Unlike prior studies that have conducted quasi-experiments for causal inference in software engineering by employing difference-in-differences~\cite{8501934,Yang2022,fang2022damn} or CausalImpact~\cite{10.1109/ICSE.2019.00059}, we are unable to employ these methods.
This is because these methods require the values of the outcome variables in the periods before and after the treatment, but data on the number of sponsors at a given point in time were not available at the time we conducted our analysis.\footnote{We contacted the GitHub team in a public forum and they responded that they would consider making the sponsor count data publicly available; we do not provide a link to that form because of the double-anonymous submission.}
Therefore, in this analysis, we apply a statistical matching method called propensity score matching (PSM), which attempts to estimate the effect of treatment by constructing a control group by matching each treated unit with a non-treated unit with similar characteristics~\cite{imbens_rubin_2015}.
PSM predicts the probability of belonging to the treatment and control groups based on observed predictors. Some of the studies mentioned above used PSM to prepare data for the treatment and control groups~\cite{8501934,fang2022damn}.

To collect developers as potential members of a control group, we contacted the authors of previous work~\cite{shimada2022github} to obtain the list of GitHub users who had participated in \GHS. From 3,697 sponsored and 5,666 non-sponsored developers collected in July 2021 for the previous study, we identified 1,930 and 4,913 developers who had not deleted their GitHub accounts and whose \GHS~profiles do not appear in our tweet data (neither in their own tweets nor in tweets from others).
Potential members of the treatment group are developers whose \GHS~profiles appear in the ``sponsor template'' tweets, that is, ``My \GHS~profile is live!''.
By targeting only ``sponsor templates'', the influence of wording differences can be eliminated. We observed that ``sponsor template'' appears most often after ``sponsor template'', which are free-text tweets (see Section \ref{sssec:rq13}).
To limit developers to those who started using \GHS~at the same period as control group developers, we collected only those developers whose \GHS~profiles appeared in such tweets by July 2021 and identified 568 developers.

The following are variables of developers used in the logistic regression to estimate the propensity score for PSM.

These variables have been used in previous related studies: for example, sponsored developers sponsor more than non-sponsored developers~\cite{shimada2022github}, sponsored developers form language-specific clusters that sponsor each other~\cite{shimada2022github}, and the number of followers is the most important feature for predicting long-term contributors~\cite{8721092}.
All values were measured in August 2022.
\begin{itemize}
    \item \textbf{repositories}: The number of public repositories created.
    \item \textbf{sponsoring}: The number of developers sponsoring.
    \item \textbf{openedPRs}: The number of opened pull requests.
    \item \textbf{reviewedPRs}: The number of reviewed pull requests.
    \item \textbf{followers}: The number of followers.
    \item \textbf{organizations}: The number of joined organizations.
    \item \textbf{language}: Categorical variable for the primary programming language determined by the method described in Section~\ref{ssec:quantitative}. The values are the top 10 languages (JavaScript, Python, PHP, C\#, Go, Java, TypeScript, C++, Ruby, and C) and others (including undetermined) seen in Table~\ref{tab:rq1.1.2}. In regression model building, dummy variables are prepared that take a value of 0 or 1 indicating the absence or presence of a particular language.
\end{itemize}

\begin{figure}
  \centering
\includegraphics[width=.7\linewidth]{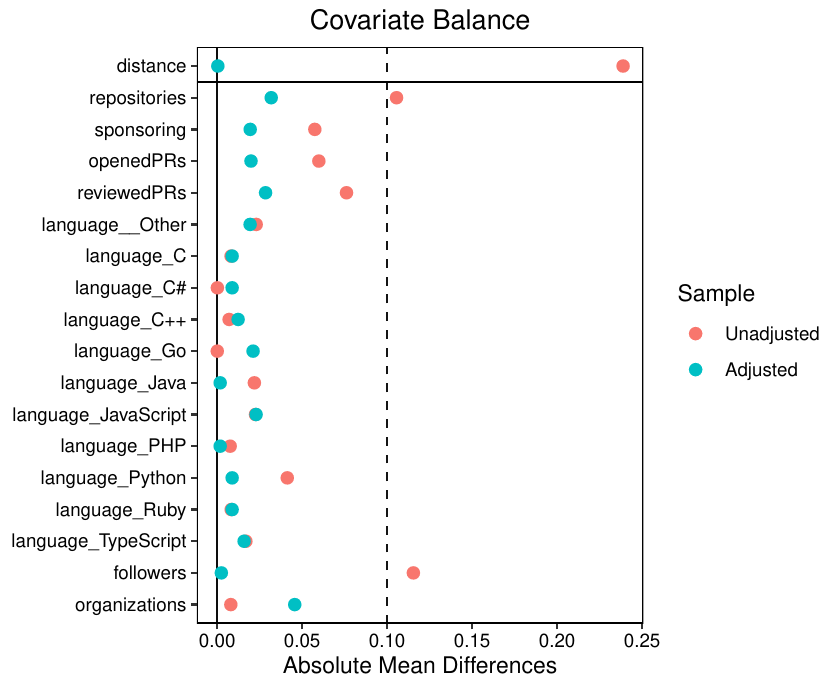}
  \caption{Covariate balance before (unadjusted) and after (adjusted) propensity score matching.}
  \label{fig:balance}
\end{figure}
We obtained 1,094 matched developers out of 7,411 ($1,930+4,913+568$) developers from the PSM.
Figure~\ref{fig:balance} shows how the absolute mean differences have decreased as a result of the matching, from \texttt{unadjusted} to \texttt{adjusted}
(unadjusted indicates all developers before matching, and adjusted indicates matched developers).
None of the absolute mean differences of \texttt{adjusted} exceeds 0.10, which means that we obtained developer matches for the treatment and control groups with a balanced distribution of covariates~\cite{imbens_rubin_2015}. This balance is a measure of the quality of the propensity score matching and we achieved the well-established and well-cited threshold~\cite{https://doi.org/10.1002/sim.3697,NORMAND2001387}.

To estimate the impact of \GHS~profile mentions in tweets, a linear regression is performed using the above variables and the variable \texttt{treatment}, which takes a value of 0 or 1 that indicates the presence or absence of \texttt{general template} tweets.
The outcome variable is the number of sponsors obtained by each developer, measured in August 2022. Therefore, this analysis estimates the impact of tweeting ``My \GHS~profile is live!'' on the number of sponsors as of August 2022, for early adopters starting \GHS~and tweeting from May 2019 (\GHS~launched) through July 2021.

\section{Results}

This section presents answers to our research questions.

\subsection{\RqOne}

The results of the analysis of the characteristics, participants, context, and timing of tweets mentioning \GHS~profiles are presented in this section.


\begin{table}[t]
\caption{Frequency of written languages of tweets that contain links to \GHS~profiles.}
\centering
\footnotesize
\label{tab:rq1.1.1}
\begin{tabular}{lcc}
\toprule
 \textbf{written languages}   & Person & Organization\\
\midrule
English      & 3,074 (94\%)& 479 (97\%)        \\
Japanese     & 151 (5\%) & 5 (1\%) \\
Undetermined & 18 (0\%) & 1 (0\%)        \\
Spanish      & 8 (0\%) & 1 (0\%)           \\
Other         & 19 (0\%)  & 4 (0\%)                \\
\midrule
sum  & 3,270 (100\%) & 496 (100\%) \\
\bottomrule
\end{tabular}
\end{table}

\begin{table}[t]
\caption{Frequency of GitHub account types and primary programming languages of distinct \GHS~profiles.}
\centering
\footnotesize
\label{tab:rq1.1.2}
\begin{tabular}{llcc}
\toprule
\multirow{1}{*}&\textbf{programming languages} & Person         & Organization \\
\midrule
&JavaScript & 816 (25\%) & 89 (18\%) \\
&Python & 333 (10\%) & 47 (9\%) \\
&PHP & 309 (9\%) & 44 (9\%) \\
&C\# & 228 (7\%) & 22 (4\%) \\
&Go & 180 (6\%) & 21 (4\%) \\
&Java & 153 (5\%) &21 (4\%) \\
&Other & 1251 (38\%)  & 252 (51\%)  \\

\midrule
&sum  & 3,270 (100\%) & 496 (100\%) \\
\bottomrule
\end{tabular}
\end{table}

\subsubsection{\textbf{RQ1.1: \RqOneDotOne}}
\label{sssec:rq11}
We investigated the written languages, GitHub account types, and the primary programming languages of the developers mentioned in the tweets. These elements were categorized based on whether they originated from personal or organizational accounts. This initial analysis serves as a foundation for our subsequent in-depth investigation, offering an initial understanding of the nature of these tweets.

\textbf{Written languages.} Table~\ref{tab:rq1.1.1} presents the frequency of written languages in tweets that contain links to \GHS~profiles.
Compared to the ranks and portions of the written languages of general tweets~\cite{hong2011language}, in English \GHS~profile tweets, both personal and organizational accounts make up a significantly larger portion than general English tweets (51\%). Japanese tweets rank second in general tweets, comprising 5\% of personal accounts and 1\% of organizational accounts. Spanish also stands out as a top contributor among the top ten languages used frequently in general tweets.

\textbf{GitHub account types.} As seen in Table~\ref{tab:rq1.1.1} and Table~\ref{tab:rq1.1.2}, most \GHS~profiles mentioned in tweets are associated with personal accounts, accounting for 87\% of 3,766 distinct \GHS~profiles. Approximately a fifth of \GHS~profiles in tweets are associated with organizational accounts, representing 13\% of distinct \GHS~profiles in the obtained tweets.
According to GitHub's advanced search engine~\cite{githubBuildSoftware}
in August 2022, 18,129 personal GitHub accounts had activated \GHS, accounting for 91\%. Furthermore, only 9\% of all GitHub accounts (1,889) that activated \GHS~are organizational accounts. Comparing \GHS~profiles that were posted on Twitter/X and all \GHS~profiles on GitHub, they tend to share a similar trend for GitHub account types.

\textbf{Programming languages.} Among the 3,766 distinct \GHS~profiles mentioned in the collected tweets, JavaScript stands out as the most prominent language, with 25\% of personal accounts, suggesting its popularity among individual users. Conversely, its relatively lower representation in organizational accounts (18\%) may indicate a preference for other languages in professional settings. Python, with 10\% of usage among personal accounts, appears to be a language of choice for individual developers, potentially due to its versatility and readability. The prevalence of Python and PHP, both of them at 9\%, among organizational accounts hints at their significance in enterprise-level development projects, as seen in Table~\ref{tab:rq1.1.2}. In the ``other'' category of coding repositories, where many instances are labeled as ``None'', there are organizations like PJSoftCo.\footnote{https://github.com/PJSoftCo} They are a prime example of how GitHub organizations are using sponsorship funds to invest in their organization-wide documentation. Comparing these results with previous work~\cite{shimada2022github}, we find that, except for \texttt{Undetermined}, the top four programming languages are exactly the same. 
The top ten primary programming languages are the same on individual \GHS~and \GHS~profiles that were posted on Twitter/X.

\begin{table}[t]
    \caption{Frequency of relationships between tweet authors and linked \GHS~profiles.}
    
    \label{tab:rq1.2}
    \centering
\footnotesize
    \begin{tabular}{lrr}
    \toprule
     & \textbf{looking for sponsors} & \textbf{sponsors} \\
    \midrule
    same & 169 (48\%) & - \\
    user & - & 55 (16\%) \\
    non-specific & 14 (4\%) & 113  (32\%) \\
    \midrule
    sum & 183 (52\%) & 168 (48\%)\\
    \bottomrule
    \end{tabular}
\end{table}

\subsubsection{\textbf{RQ1.2: \RqOneDotTwo}} Table~\ref{tab:rq1.2} shows the results of the coding for \textbf{RQ1.2}. As mentioned in Section~\ref{ssec:qualitative}, we separated tweets by purpose, distinguishing developers who mention their \GHS~profiles to look for sponsors from those who are sponsors. Developers that were looking for sponsors mentioning their own \GHS~profiles in tweets is the most frequently occurring case, accounting for 48\% of the sample. However, it is also common that sponsors mention \GHS~profiles of other GitHub accounts, accounting for 32\% of the sample. We observe that sponsors also explicitly mentioned \GHS~profiles of others due to dependencies or other benefits, accounting for 16\% of the sample. In previous work~\cite{shimada2022github}, Shimada et al. showed that developers sponsoring others via \GHS~due to dependencies is the most frequent reason for sponsoring. In the context of Twitter/X, our result partially agrees with their observations.

\begin{table}
    \caption{Frequency of context of the \GHS~profile mentions on Twitter/X.}
    \label{tab:rq1.3}
\footnotesize
    \centering
    \begin{tabular}{lrr}
    \toprule
     & \textbf{looking for sponsors} & \textbf{sponsors} \\
    \midrule
    generic advertisement &  72 (21\%) & - \\
    donation appreciation & 34 (10\%) & - \\
    sponsor template &  33 (9\%) & - \\
    advertisement of developer &  9 (3\%) &  7 (2\%) \\
    advertisement with new functionality &  9 (3\%) & - \\
    advertisement with new information &  8 (2\%) & - \\
    sustainability & 5 (1\%) & - \\
    advertisement with early access &  5 (1\%) & -\\
    income & 2 (1\%) & -\\
    advertisement with event &  3 (1\%) & - \\
    set example / peer pressure &  2 (1\%) &  4 (1\%) \\
    advertisement of organization & 1 (0\%) &  2 (1\%) \\
    donation to developer announcement & - &  101 (29\%) \\
    donation to organization announcement & - &  33 (9\%) \\
    donation to developer template & - & 16 (5\%) \\
    donation to organization template & - & 5 (1\%) \\
        \midrule
        sum & 183 (52\%) & 168 (48\%)\\
        \bottomrule
        \end{tabular}
\end{table}

\begin{table}[ht]
    \caption{Frequency of timing of tweets that contain links to \GHS~profiles.}
    \label{tab:rq1.4}
\footnotesize
    \centering
\begin{tabular}{lrr}
    \toprule
     & \textbf{looking for sponsors} & \textbf{sponsors} \\
    \midrule
    start &  62 (18\%) & -\\
    no specific timing &  46 (13\%) & 111 (32\%) \\
    donation &  35 (10\%) & - \\
    update &  20 (6\%) & - \\
    reach goal & 8 (2\%) & - \\
    release & 6 (2\%) & - \\
    event & 3 (1\%) & - \\
    resignation / paycut &  3 (1\%) & - \\
    benefit & - &  55 (16\%) \\
    activity spike & - &  2 (1\%) \\
        \midrule
        sum & 182 (52\%) & 169 (48\%)\\
        \bottomrule
        \end{tabular}
\end{table}

\subsubsection{\textbf{RQ1.3: \RqOneDotThree}}
\label{sssec:rq13}
In Table~\ref{tab:rq1.3}, the frequency of various \GHS~profile mentions on Twitter/X is presented. Most sponsors mentioned \GHS~profiles on Twitter/X in the context of donating to personal or organizational GitHub accounts, accounting for 29\% and  9\%, respectively. Sponsors also mentioned their donation to personal or organizational GitHub accounts using GitHub's tweet templates, accounting for 5\% and 1\%, respectively. 

In addition to donations, developers looking for sponsors mentioned \GHS~profiles on Twitter/X to advertise their own profile (21\%) or to advertise profiles of other personal GitHub accounts (3\%).
Specifically, developers looking for sponsors advertise their own profile with updates on the functionality of the project (3\%), updates on the profile (2\%), early access features (1\%), and events (1\%). As with the donation, developers looking for sponsors advertised their own profiles using GitHub's tweet templates, accounting for  9\%. Furthermore, a few developers looking for sponsors also posted tweets to encourage others to donate by setting an example or applying peer pressure, accounting for 1\%. 

Some developers looking for sponsors use Twitter/X as a channel to express appreciation to sponsors (10\%). Furthermore, a few tweets from developers looking for sponsors mention \GHS~profiles in the context of sustainability of the project or the financial income of developers. In particular, we see that several tweets are posted to share \GHS~updates in the context of the \texttt{Log4j} vulnerability~\cite{mitreCVE202144228}
that was exploited in December 2021. For example, ``\textit{It's nice to see that a month after the Log4Shell vulnerability Log4j's maintainer has 101 \GHS~instead of 3, including corporate accounts such as Amazon Web Services}''.
\subsubsection{\textbf{RQ1.4: \RqOneDotFour}}\label{sec:4.1.4}

Table~\ref{tab:rq1.4} presents the frequency of different types of timing when different types of developers mentioned \GHS~profiles on Twitter/X. Regardless of the different types of developers, we find that most of the tweets (45\%) do not specify an explicit explanation of the reason for the tweet's posting at that particular time. However, we can see that some tweets (16\%) were posted at a time when sponsors benefited from a project. Developers that were looking for sponsors posted those tweets during the initiation of \GHS~profiles (18\%), at the time of donation (10\%), or when updating projects or profiles (6\%). Furthermore, some developers that were looking for sponsors posted tweets with \GHS~profile mentions when they need financial resources due to changes in their work arrangements.

As seen in Table~\ref{tab:rq1.4}, we found an interesting type of timing with regard to when \GHS~profiles were mentioned in tweets: activity spikes, i.e., a sponsor donated due to an activity spike of a developer. Inspired by this code, we conducted a quantitative study to analyze the correlations between contributions of developers and \GHS~profile mentions on Twitter/X. 
Table~\ref{tab:rq1.4.1} presents comparisons among three periods across a set of GitHub contribution types. 
Since the GitHub organization account lacks information on activity, we excluded tweets that contain \GHS~from organizations. Then, we focus on tweets from distinct developers that tweeted with ``My \GHS~profile is live!''. Only 810 tweets were included out of the 10,531 English tweets. The rationale behind this choice was to specifically analyze the initial reactions and sentiments expressed by users who had just enabled their \GHS~account. Our primary goal was to capture the immediate activity of individuals in this specific context. Our sample is representative of \GHS~users' initial tweets about their GitHub Sponsors account, but not of all tweets in our dataset. We observe that most of the mean values for the week in which a tweet was posted are higher than the corresponding values in the week before or after. 
For the contribution activities of Opening discussion, Committing, and Creating repository, there are significant differences between a week before posting this tweet and the week when the tweet was posted, and between the week when the tweet was posted and a week after posting this tweet, with at least negligible effect sizes (Committing shows small effect size). In addition to these activities, comparing a week before posting the tweet and the week when the tweet was posted, we find that developers proposed significantly more pull requests, with negligible effect size. These results indicate that when developers posted their \GHS~profile on Twitter/X, they generally contribute more actively to OSS projects.

\begin{tcolorbox}
\textbf{Summary}: 
Of the \GHS~profiles mentioned in the tweets, 87\% belong to individual developers, whose top primary languages were JavaScript, Python, and PHP.
Such tweets were posted by the owners of the profiles or by others who depended on the work of the developer they sponsored. Developers looking for sponsors were more active on GitHub during the week in which tweets linking to their \GHS~profile were posted. 
\end{tcolorbox}

\begin{table}[htp]
    \caption{Comparisons among three periods of GitHub contributions}
    \label{tab:rq1.4.1}
    \footnotesize
    \setlength{\tabcolsep}{1pt} 
    
    \begin{threeparttable}
        \begin{tabular}{lccc|ccc|ccc}
            \toprule
            & \multicolumn{3}{c}{\textbf{Before}} & \multicolumn{3}{c}{\textbf{During}} & \multicolumn{3}{c}{\textbf{After}} \\
            \cmidrule{2-10}
            & mean & Q3 & effect size &  mean & Q3  &  effect size & mean & Q3 &  effect size \\
            \midrule
            Opening PR & 1.12 & 1 & \cellcolor[HTML]{D3D3D3}\textbf{0.0929***} & \textbf{1.75} & \textbf{2} & - & 1.5 & 1 & - \\
            Subm. PR review & 1.2 & 0 & - & \textbf{1.28} & 0 & - & 1.19 & 0 & - \\
            Opening issue & 0.61 & 0 & - & \textbf{1.01} & 1 & - & 0.84 & 1 & - \\
            Opening disc. & 0.03 & 0 & \cellcolor[HTML]{D3D3D3}\textbf{0.0419***} & \textbf{0.12} & 0 & - & 0.06 & 0 & \cellcolor[HTML]{D3D3D3}\textbf{0.0274**} \\
            Answering disc. & 0.02 & 0 & - & 0.03 & 0 & - & \textbf{0.04} & 0 & - \\
            Committing & 13.49 & 15 & \cellcolor[HTML]{A9A9A9}\textbf{0.181***} & \textbf{18.31} & \textbf{21} & - & 14.4 & 16.75 & \cellcolor[HTML]{A9A9A9}\textbf{0.153***} \\
            Contr. to priv. repo. & 7.69 & 6 & - & \textbf{8.39} & \textbf{7} & - & 7.42 & 5.75 & - \\ 
            Creating repo. & 0.52 & 0 & \cellcolor[HTML]{D3D3D3}\textbf{0.0938***} & \textbf{0.71} & \textbf{1} & - & 0.5 & 0 & \cellcolor[HTML]{D3D3D3}\textbf{0.103***} \\
            Joining org. & 0.01 & 0 & - & 0.01 & 0 & - & 0.01 & 0 & - \\
            \bottomrule
        \end{tabular}
        \begin{tablenotes}
           \footnotesize
        \item  * p-value $<$ 0.05; ** p-value $<$ 0.01; and *** p-value $<$ 0.001. The Cliff's delta effect size with thresholds~\cite{romano2006appropriate} are highlighted in  \sethlcolor{Negligible}\hl{Negligible} \sethlcolor{Small}\hl{Small} \sethlcolor{Medium}\hl{Medium} \sethlcolor{Large}\hl{Large}. The hyphen (-) is used as a placeholder for cases where p-value $\geq$ 0.05 indicates there is no significant difference in the comparison or when comparing to itself. 
        \end{tablenotes}
    \end{threeparttable}
\end{table}

\begin{table}[htbp]
    \centering
    \caption{Frequency of response to tweets.}
    \label{tab:rq2.2}
    \footnotesize
    \begin{tabular}{lrr}
      \toprule
      & \textbf{looking for sponsors} & \textbf{sponsors} \\
      \midrule
      other &  165 (47\%) & 138 (39\%) \\
      appreciation of work &  9 (3\%) & - \\
      support &  7 (2\%) &  3 (1\%) \\
      appreciation of donation & 2 (1\%) & 25 (7\%) \\
      emoji only & - & 2 (1\%) \\
      \midrule
      sum &  183 (52\%) & 168 (48\%)\\
      \bottomrule
    \end{tabular}
\end{table}

\begin{table*}
    \caption{Comparisons among donation and crowd-funding platforms in Twitter interactions.}
    \label{tab:rq2.1}
\footnotesize
    \centering
    \begin{threeparttable}
    \begin{tabular}{lrrrrrrr}
    \toprule
    & \multicolumn{2}{c}{\textbf{like}} & \multicolumn{2}{c}{\textbf{retweet}} & \multicolumn{2}{c}{\textbf{reply}} &  \multirow{2}{*}{\#}\\ \cline{2-7}
    & median & effect size & median & effect size & median & effect size & \\
    \midrule
Open Collective & 4 & - & 1  & \cellcolor[HTML]{A9A9A9}\textbf{0.216***} & 0 &  - & 88   \\
Patreon & \textbf{6} & \cellcolor[HTML]{A9A9A9}\textbf{0.265***} & \textbf{2} &   \cellcolor[HTML]{A9A9A9}\textbf{0.278***} & 0 &  \cellcolor[HTML]{D3D3D3}\textbf{0.132***} & 228  \\
\GHS~& 3 & -  & 0 &  - & 0 &  - & 10,440 \\
    \bottomrule
    \end{tabular}
\begin{tablenotes}
        \footnotesize
        \item * p-value $<$ 0.05; ** p-value $<$ 0.01; and *** p-value $<$ 0.001. The Cliff's delta effect size with thresholds~\cite{romano2006appropriate} are highlighted in  \sethlcolor{Negligible}\hl{Negligible} \sethlcolor{Small}\hl{Small} \sethlcolor{Medium}\hl{Medium} \sethlcolor{Large}\hl{Large}. The hyphen (-) is used as a placeholder for cases where p-value $\geq$ 0.05 indicates there is no significant difference in the comparison or when comparing to itself (\GHS).
      \end{tablenotes}
      \end{threeparttable}
\end{table*}

\begin{table}[tbp]
    \caption{Frequency of response to tweets.}
    \label{tab:rq2.2}
\footnotesize
    \centering
    \begin{tabular}{lrr}
    \toprule
     & \textbf{looking for sponsors} & \textbf{sponsors} \\
    \midrule
none & 140 (40\%) & 130 (37\%) \\
other & 27 (8\%) & 9 (3\%) \\
support & 6 (2\%) & 2 (1\%) \\
appreciation of work & 6 (2\%) & - \\
appreciation of donation & 2 (1\%) & 25 (7\%) \\
emoji only & - & 2 (1\%) \\
    \midrule
    sum & 182 (52\%) & 169 (48\%)\\
    \bottomrule
    \end{tabular}
\end{table}

\subsection{\RqTwo}
\subsubsection{\textbf{RQ2.1: \RqTwoDotOne}}


\revise{Table~\ref{tab:rq2.1} presents comparisons between donation and crowd-funding platforms in Twitter interactions.} For median values, Patreon tweets received the highest number of likes, accounting for six. Additionally, Patreon tweets were retweeted twice, which is the highest number of retweets in terms of median values. The median values of replies for the three platforms are zero. In terms of likes, the p-value of Patreon vs.~\GHS~is less than 0.05 and Cliff's delta is 0.265, indicating Patreon and \GHS~have a significant difference with a small effect size. In terms of retweets, Open Collective and \GHS~have a significant difference (i.e., p-value $<$ 0.05) with a small effect size (i.e., 0.147 $\leq$~$|$\textit{delta}$|$ $<$ 0.33). Comparing Patreon and \GHS~in terms of retweets, there is a significant difference with a small effect size. In terms of replies, we find that Patreon and \GHS~have a significant difference with a negligible effect size (i.e., Cliff's delta is 0.132). In conclusion, while the number of tweets containing \GHS~profile mentions is much larger, tweets that link to Patreon or Open Collective in the context of OSS receive more likes and retweets.



\subsubsection{\textbf{RQ2.2: \RqTwoDotTwo}}
Table~\ref{tab:rq2.2} shows the results of our coding of replies to tweets that mentioned \GHS~profiles. 
We can see that most tweets (86\%) do not receive a response on Twitter/X. For the remaining 14\%, the majority consists of expressions of appreciation for donations (8\%).
Since Twitter/X is an informal communication channel, we observe that some responses consist only of one or more emoji.

\begin{table}[htp]
    \centering
    \caption{Causal inference impact of \GHS~Profile Mentions in ``My GitHub Sponsors Profile is Live!'' Tweets.}
    \label{tab:rq2.3}
\footnotesize
    \begin{tabular}{llll}
    \toprule
    & \textbf{estimate} & \textbf{std. error} & \textbf{p} \\
    \midrule
treatment       &    \textbf{1.22}     & 0.452        &  \textbf{0.00681} \\
repositories    &   -0.000818 & 0.00209     & 0.696 \\
sponsoring      &    1.12     & 0.194        &  1.01e-8 \\
openedPRs       &    0.000432 & 0.000457     & 0.345 \\
reviewedPRs     &    0.00301  & 0.00140      &  0.0325 \\
followers       &    0.00271  & 0.000345     &  8.42e-15 \\
organizations   &   -0.0637   & 0.0870      & 0.465 \\
language\_JavaScript & -1.45  &    0.657       &  0.0279 \\
language\_Python    & -1.18   &   0.856       &  0.168 \\
language\_PHP   &     -0.343  &   0.906       & 0.705 \\
language\_C\#   &       0.559 &    0.973        & 0.566 \\
language\_Go    &     -0.933  &   1.02        & 0.360 \\
language\_Java  &     -1.66   &   1.15        &  0.149 \\
language\_TypeScript & -0.692 &    1.04        & 0.505 \\
language\_C++   &     -1.07   &   1.44        & 0.458 \\
language\_Ruby  &     -1.01   &   1.50        & 0.500 \\
language\_C     &     -0.659  &   1.21        & 0.586 \\
    \bottomrule
    \end{tabular}
\end{table}

\subsubsection{\textbf{RQ2.3: \RqTwoDotThree}}
\label{sssec:rq23}
 Table~\ref{tab:rq2.3} summarizes the regression result. As seen in the coefficient estimate of \texttt{treatment}, there is a statistically significant positive effect of \GHS~profile mentions in tweets on the number of sponsors acquired.
As the average of the expected causal effect of treatment on individuals in the treatment group, called Average Treatment Effects on the Treated (ATT), we find that tweet mentions have an impact of increasing the number of sponsors by 1.22.
However, note that the medians, Q3, and means for the matched treatment and control groups are \{$0, 2.00, 2.56$\} and \{$0, 1.00, 1.30$\}, respectively, indicating a skewness in the developers who obtained sponsorship, that is, the effects are not uniform.
\begin{tcolorbox}
\textbf{Summary}: 
\GHS~profile mentions have a positive impact on the number of sponsors acquired, increasing the number of sponsors by 1.22. On Twitter/X, tweets mentioning \GHS~receive fewer interactions than those mentioning Patreon or Open Collective, and most tweets do not attract replies.
\end{tcolorbox}

\section{Threats to Validity}
\textit{Subjective nature of coding.} 
We conducted qualitative analyses of a statistically representative sample of tweets. 
The codes we assigned to different tweets may be inadequate due to the subjective nature of understanding the various coding schemata. To migrate this threat, we require kappa agreements of at least ``substantial agreement'' to ensure a common understanding of the coding schemata among all four annotators. Then, we initiated another round of coding between the first two authors for the remaining sample. By recalculating kappa agreements, we can see the improvement in understanding the various coding schemata. For example, Cohen's kappa increased from 0.62 for the first 30 tweets to 0.85 for the reaming 321 tweets in the coding of the timing of tweets. The final results are based on the codes on which the authors, after discussion, reached a consensus and collectively agreed.
%

\textit{Limitations in causal inference result.}
We compared developers with and without tweets who started GitHub Sponsors in the same period and engaged in similar activities, but we may have missed important developer characteristics other than the metrics we measured.
The result is best interpreted as an increase in sponsors acquired through social activities on Twitter/X, rather than simply tweeting ``My GitHub Sponsors profile is live!''.
In this study, we only analyzed the impact of tweets using such a template, so the impact of free-text tweets is unknown.
In addition, since this analysis was conducted on early adopters, it is not possible to generalize whether similar effects will be seen in the future, so a continued analysis is needed.

\textit{Multiple \GHS~profiles in the same tweet.} There is a small number of cases where the same tweet contains multiple \GHS~profile links to different GitHub accounts. Since these cases are rare (i.e., only five tweets) and to avoid confusion in our analyses, we exclude these tweets from our analyses.

\textit{Only tweets with \GHS~profiles links. Simple keyword searches would have introduced too much noise to our large-scale analysis. 
To avoid false positives, we only recovered tweets with \GHS~profile links, but we acknowledge that other relevant tweets without links may have been omitted.}

\textit{Primary programming languages of the developers.} 
We considered the most common primary language of the repositories to which each developer contributed as the primary language of that developer. This means it could happen for some users, for example, that the most common language of a developer's contributed repositories is Java, but the developer may only contribute the documentation of these Java repositories, whereas committing Python code to another project. Therefore, it is important to acknowledge this potential limitation in accurately capturing a developer's primary programming language through this methodology.

\textit{The number of tweets mentioning GitHub Sponsors has a different scale of data compared with other sponsorship platforms.} We collected  10,440 tweets for GitHub Sponsors compared to other platforms: 4 for PayPal, 88 for Open Collective, and 228 for Patreon. It is important to recognize that this difference in data size may affect the robustness and generalizability of our conclusions.

\textit{External validity is concerned with our ability to generalize based on our results.} In Section~\ref{sec:4.1.4}, we used a subset of 810 tweets from a pool of 10,531 English tweets. It is crucial to acknowledge that the chosen subset may not fully represent the broader spectrum of reactions across all types of tweets related to \GHS. Users who express their thoughts in different formats or use alternative phrases may not be fully captured in our analysis. 

\section{Discussion}

This section presents implications and future work from this study.

\textbf{Implications.} We categorized the practical implications for diverse groups of individuals by offering tailored guidance and recommendations that align with the specific concerns and interests of each stakeholder group.

\begin{itemize}[topsep=1pt, partopsep=1pt, itemsep=1pt, parsep=1pt]
 
    \item \textit{Developers seeking sponsorship}: our study shows that mentioning \GHS~profiles in tweets has a positive impact on the number of sponsors acquired. 
    The finding that the number of sponsors acquired increased depending on whether they tweeted, is evidence of the importance of social media and should encourage developers to go beyond the GitHub platform in order to attract sponsorship.
    Additionally, our research reveals many different types of messages surrounding \GHS~in various contexts, providing insights that might assist others in crafting their own effective social media strategies for sponsorship engagement.

    \item \textit{Developers interested in sponsoring}: Within our sample, approximately half of the participants are sponsors. This finding underscores the importance of encouraging users who depend on OSS projects to actively promote the developers they rely on, even if the sponsorship amount is not substantial. 
    Engaging in social media promotion can significantly enhance the visibility of these developers, allowing their exceptional work to reach a wider audience and garner increased recognition.



    \item \textit{Companies}: The relationship between companies and OSS projects is undergoing a pivotal change, largely driven by an expanding recognition of sustainability issues inherent in OSS. 
    Instead of merely expressing dissatisfaction with the lack of sustainability in these projects, our study offers evidence that a two-pronged approach of corporate sponsorship and active social media engagement could be an effective strategy for businesses.
    This strategy allows them to constructively engage with OSS projects they rely upon, particularly those struggling with sustainability, thereby addressing their concerns and contributing to potential solutions.
\end{itemize}
\textbf{Future Work.} As our study is positioned as an early adopter study, we have not yet obtained conclusive evidence of a significant impact of financial support on OSS sustainability at this stage. 
Therefore, 
further investigation of the potential impact of financial support on sustainability is needed.

We have focused on Twitter/X as the starting point for our exploration. For future research, there is significant value in extending our analysis to encompass posts from multiple social media platforms (e.g., Facebook, Reddit) to gain a more comprehensive understanding of these dynamics.

As part of our investigation into \textbf{RQ2.1}, we found that tweets linking exclusively to \GHS~were more common, and among those platforms, GitHub is the only one that provides Twitter/X templates for developers looking for sponsors and Twitter/X templates for sponsors. However, \GHS~received fewer responses compared to tweets promoting alternative sponsorship platforms. Since \GHS~launched 4--6 years later than Open Collective and Patreon, so it had less time to solidify its position and gain widespread awareness. 
Further research is needed to determine the importance of the template if such a sponsorship platform provides a template when users are trying to advertise on social media, and strategies for increasing the response and engagement of tweets containing only links to \GHS.

In light of our findings regarding the skewness in the developers who obtained sponsorship (Section \ref{sssec:rq23}),
factors other than tweets may play an important role in sponsor acquisition. Thus, 
further research is needed to explore and identify these additional factors that contribute to the sponsor acquisition process.

In subsequent studies, it would be valuable to investigate the specific strategies and practices employed by organizations when leveraging social media platforms to disseminate project updates.
By examining the relationship between these practices and the resulting engagement levels within the community, we can gain insights into the effectiveness of such approaches and their potential for enhancing community involvement. Moreover, it is worth considering a more in-depth investigation into how the domain and functionalities of open-source projects can impact and guide the dynamics of sponsorship, such as evaluations of a project's sustainability and its sponsorship status~\cite{6127835}.

\section{Related Work}

In this section, we situate our work with respect to the literature on donations and one potential advertising channel, Social Media.

\textbf{Donation.}
OSS development heavily relies on volunteer contributions, as highlighted in a recent GitHub survey~\cite{github2017survey}, revealing that just 23\% of respondents contribute to open source as part of their job description.
Despite more employees being paid for contributing to OSS projects during work hours~\cite{riehle2014paid}, developers still perceive compensation asymmetry in OSS projects~\cite{atiq2016impact}.
OSS projects that are distributed unequally may fail if they are mismanaged and financial benefits are a factor in the sustainability of OSS projects~\cite{atiq2016impact}.
Donation is one of the common ways to obtain these financial benefits~\cite{eghbal2019handy} to support OSS projects, in addition to Bounty~\cite{eghbal2019handy}. In a mixed-method empirical study, Overney et al.~\cite{10.1145/3377811.3380410} found that only a few projects (0.04--0.2\%) ask for donations, primarily using platforms like PayPal and Patreon. These projects tend to be more active, more mature, and more popular.


Recently, Zhou et al.~\cite{zhou2022studying} explored donations on the \texttt{Open}\\ \texttt{Collective} platform that support open-source projects. They indicated the influence of individual donors; although corporate donors tend to donate more money than individual donors for an individual donation, the total donation amount from individual donors is greater than corporate donors. However, corporate collectives are more likely to receive a larger total donation amount
than individual collectives.
Regarding the study on \GHS, Shimada et al.~\citep{shimada2022github} revealed that developers typically do not have channels at their disposal to attract sponsors and communicate with those who might be interested in donating.
Zhang et al.~\cite{zhang2022and} discovered that sponsorship through \GHS~has a short-term impact on developers' activities. Their survey highlighted key challenges, including the difficulty of attracting sponsorship and the absence of corporate support.

\textbf{Social Media.}
Social media channels are one way of communicating and advertising in the world of developers. Different social media channels play different roles and have different impact on OSS projects, e.g., facilitating communication~\cite{black2010survey, storey2010impact}, awareness of the status of other developers~\cite{begel2010social, black2010using}, gaining attention~\cite{borges2019developers, maldeniya2020herding, trockman2018adding}, and attracting new contributors~\cite{blincoe2016understanding, maldeniya2020herding,qiu2019signals}.

Researchers studied the use of microblog services such as Twitter/X in software development~\cite{borges2019developers}. The tweets of developers differ from those of the general public in terms of length, use of URLs, and \texttt{@-mentions}, and software microbloggers are more tightly
knit than general microbloggers~\cite{bougie2011towards, tian2014exploratory}. Twitter/X is widely adopted in the software engineering community~\cite{black2010survey, storey2016social}. 
Tian et al.~\cite{tian2012does} found that knowledge sharing, technical discussion, and software product updates are the most frequent categories of developers' tweets. 
Fang et al.~\cite{fang2020need} proposed an approach to cross-link users on Twitter/X and GitHub; they observed that tweeting patterns appear in tweets from different developer roles when including GitHub links in their tweets (e.g., repository owners prompt their projects instead of discussing specific software artifacts).

For the impact of Twitter/X, Singer et al.~\cite{singer2014software} indicated that Twitter/X can help developers become aware of industry changes, learn, and build work relationships in communities. Mezouar et al.~\cite{mezouar2018tweets} found that tweets from end users can lead
to early discovery of bugs in web browsers. Fang et al.~\cite{fang2022damn} explored the causal effects of Twitter/X on the attraction of stars and contributors by analyzing tweets that contain links to GitHub repositories. They found that Twitter/X has a statistically significant and sizeable effect to help make projects popular (i.e., stars) but only a small effect to attract new contributors (i.e., commits). Moreover, these newly attracted contributors showed to be more active in OSS projects when they had prior Twitter/X interactions with the tweet authors.
\section{Conclusion}

There are several platforms that enable contributions to open-source software developers, but attracting sponsors in order to ensure project sustainability remains a challenge. To understand the impact of Twitter/X on helping OSS developers attract sponsors, we conducted quantitative and qualitative analyses of more than 10,000 tweets containing links to \GHS~profiles. 
We find that such tweets have a significant positive effect on the acquisition of sponsors, and that developers contribute more OSS work than usual to attract potential sponsors during the week in which they posted tweets that link to their own \GHS~profile.
%

Open-source developers who maintain an active presence on social media can attract donations that help sustain their projects. 
Our findings suggest that social media channels and donation channels are linked in the social programmer ecosystem and will continue to grow in importance for the sustainability of open source software. 

\section*{Data Availability}
\revise{The replication package includes scripts and data set, which is available at  \url{https://doi.org/10.5281/zenodo.10461383} and \url{https://github.com/NAIST-SE/GHSponsorsX}}

\section*{ACKNOWLEDGEMENTS}
\revise{This work was supported by JSPS KAKENHI Grant Numbers
JP20H00587 and JP20H05706, JSPS Grant-in-Aid for JSPS Fellows JP23KJ1589, and JST PRESTO Grant Number JPMJPR22P6.}

\bibliographystyle{ACM-Reference-Format}
\bibliography{base}

\appendix

\end{document}